\documentstyle[12pt]{article}
\begin{document}
\baselineskip=24pt
%
\title{
\vspace{-2.0cm}
\begin{flushright}
{\normalsize UTHEP-281}\\
\vspace{-0.3cm}
{\normalsize July 1994 }\\
\end{flushright}
\vspace*{2.0cm}
{\Large Pion-Nucleon Sigma Term in Lattice QCD\vspace*{0.5cm}}
\author{ M. Fukugita$^{a)\dagger}$, Y. Kuramashi$^{b)}$
         M. Okawa$^{b)}$, A. Ukawa$^{c)}$\\
\\ \\
   {\it     Yukawa Institute, Kyoto University$^{\:a)}$ }\\
   {\it     Kyoto 606, Japan }\\ \\
   {\it     National Laboratory for High Energy Physics(KEK)$^{\:b)}$ }\\
   {\it     Tsukuba, Ibaraki 305, Japan }\\ \\
   {\it     Institute of Physics,  University of Tsukuba$^{\:c)}$ }\\
   {\it     Tsukuba, Ibaraki 305, Japan }\\ \\
}}

\date{}
\maketitle
\vspace*{0.5cm}

\begin{abstract}
\baselineskip=24pt

We calculate both connected and disconnected contribution to the $\pi$-$N$
$\sigma$-term in quenched lattice QCD with Wilson quark action on a
$12^3\times 20$ lattice at $\beta=5.7$.  The latter is evaluated with the aid
of
the variant wall source method, which was previously applied
successfully for extraction of $\pi$-$\pi$ scattering lengths and
$\eta^\prime$
meson mass.  We found that the disconnected contribution is about
twice larger than the connected one.  The value for the full $\pi$-$N$
$\sigma$
term $\sigma=40-60$ MeV is consistent with the experimental estimates.  The
nucleon matrix element of the strange quark density $\bar s s$ is fairly large
in our result. \end{abstract} \vfill

\noindent ${}^\dagger$ Also at {\it Institute for
Advanced Study, Princeton, NJ 08540, U. S. A.}

\newpage

Study of the $\pi$-$N$ $\sigma$-term defined as the nucleon matrix element of
the scalar density,
\begin{equation}
{\sigma}=\frac{m_{u}+m_{d}}{2}<N|{\bar u}u+{\bar d}{d}|N>,
\label{eq:sigmaterm}
\end{equation}
has a long history and still shows some controversy.  Current
algebra and PCAC relates  the $\sigma$ term to the quantity
$\Sigma=f_\pi^2\bar
D^+(2m_\pi^2)$ where $\bar D^+(2m_\pi^2)$ is the crossing even $\pi$-$N$
scattering amplitude at the Cheng-Dashen point $t=2\mu^2$, and a dispersion
analysis of $\pi$-$N$ scattering leads to the value
$\Sigma=64(8)$MeV\cite{koch}
MeV.  On the other hand, treating the flavor symmetry
breaking to first order, one finds  $\sigma=\sigma_0/(1-y)$ with
$\sigma_0\approx 25$ MeV and $y=2<N|\bar s s|N>/<N|{\bar u}u+{\bar
d}{d}|N>$\cite{langacker}.  The two estimates cannot be reconciled unless the
nucleon
has an unexpectedly large strangeness content $y\approx 0.6$.  A possible
resolution is
that the variation of the $\pi$-$N$ amplitude from the Cheng-Dashen point
$t=2\mu^2$ to
$t=0$, where the $\sigma$ term (\ref{eq:sigmaterm}) is defined, is
substantial,
resulting in a smaller value $\sigma\approx 45$MeV\cite{update}.  Combined
with the
suggestion\cite{gasser} that one-loop chiral perturbative
corrections raises the value of $\sigma_0$ to $\sigma_0=35$MeV, this would
imply a more reasonable value $y\approx 0.2$ of the nucleon strangeness
content.

In principle lattice QCD provides a more direct  and more satisfactory
route for resolving the issue through a numerical calculation of the $\sigma$
term matrix element.  A serious technical obstacle, however, has been that
evaluation of the disconnected amplitude of quark loops and nucleon
propagators
projected onto the zero  momentum state shown in Fig.~1(a) would require a
prohibitively large number of quark matrix inversions if one were to carry it
out with the conventional method of point source.
In this article we show that the variant of the method of wall source without
gauge fixing, which has been successfully applied for calculation of
$\pi$-$\pi$
scattering lengths\cite{pipisct} and $\eta^\prime$ meson mass\cite{etamass},
also
works well for the disconnected contribution to the $\sigma$ term.  Employing
the
Wilson quark action, we find an encouraging result, $\sigma=43(7)$MeV, from an
estimate of the matrix element and the up and down quark masses, or
$\sigma=58(11)$MeV from that of $m_N\sigma/m_\pi^2$, albeit within quenched
QCD
at a fairly strong coupling of $\beta=5.7$ with the lattice spacing $a\approx
0.14$fm.

Let us note that previous estimates of the connected
contribution in quenched QCD yielded values in the range $15-25$
MeV\cite{threepoint}.  The derivative
$m_qdm_N/dm_q$ in quenched QCD, equal to the $\sigma$ term via the
Hellmann-Feynman theorem,  also gave similar values\cite{dmndmq}.  These
results were considered reasonable; sea quarks are absent in the quenched
approximation, and hence one expects $\sigma_{connected}\approx
\sigma_0=25$MeV.  This already suggested that the disconnected contribution is
substantial.  Indeed we find
$\sigma_{disconnected}/\sigma_{connected}=2.23(52)$.   Previous  attempts for
estimating  the disconnected contribution in full QCD, made either indirectly
through a comparison of the slope $m_qdm_N/dm_q$ and the connected
contribution\cite{gupta} or through the noisy estimator for quark
loop\cite{mtc}, indicated that the disconnected contribution is as large as
the
connected one.  However, both results suffered from significant  errors so
that
reaching a reliable conclusion was difficult.  We should add that the large
disconnected contribution poses a problem regarding the strangeness content,
which we shall comment upon toward the end of this article.

To extract the nucleon matrix element of the scalar density $S=\bar uu+\bar
dd$
we calculate the ratio of the three-point
function of the nucleon and scalar density to the nucleon two-point function,
each projected onto  the zero momentum state{\cite{threepoint}},
\begin{equation}
R(t)=\frac{<N(t)\sum_{t^\prime\neq 0}S(t^{\prime})\bar N(0)>}{<N(t)\bar N(0)>}
\stackrel{{\rm large}\; t}{\longrightarrow} {\rm const.}+Z_S^{-1}<N|
{\bar u}u+{\bar d}d|N>t,
\label{eq:ratio}
\end{equation}
with $Z_S$ the lattice renormalization factor for the scalar density. The
$t=0$
time slice is excluded from the sum in the numerator since we fix
the $t=0$ time slice of gauge configurations to the Coulomb gauge
to enhance nucleon signals. This procedure affects only the constant term in
({\ref{eq:ratio}}).          The connected amplitude (Fig.~1(b)) can be
calculated by the conventional source method\cite{source}.  To handle the
disconnected piece (Fig.~1(a)) we prepare a quark propagator evaluated with
unit
source at every space-time site ( except for  the $t=0$ time
slice ) without gauge fixing\cite{etamass};
\begin{equation}
G({\bf n}^{\prime},t^{\prime})=
\sum_{({\bf n}^{\prime\prime},t^{\prime\prime}\neq 0)}
G({\bf n}^{\prime},t^{\prime} ;{\bf n}^{\prime\prime},t^{\prime\prime}).
\label{eq:stwall}
\end{equation}
The product of the nucleon propagator and $\sum_{({\bf n}^{\prime},
t^{\prime}\neq 0)}{\rm Tr}[G({\bf n}^{\prime},t^{\prime})]$ equals the
disconnected amplitude up to gauge-variant non-local terms which cancel out in
the average over gauge configurations. The superior feature of this method
is that it requires only two quark matrix inversions for each gauge
configurations.  We note that with the exclusion of the $t^{\prime\prime}=0$
term in (\ref{eq:stwall}), the Fierz mixing of the quark propagator
(\ref{eq:stwall}) for the disconnected loop and nucleon valence quark
propagators is automatically avoided.

We have applied the method for the Wilson quark action in quenched QCD at
$\beta=5.7$ on a $12^3\times 20$ lattice.  We analyzed 300 configurations for
the
hopping parameter $K=0.160$, $0.164$ and 400
configurations for $0.1665$, generated with the single plaquette action
separated by 1000 pseudo-heat bath sweeps.  In order to avoid
contaminations from the negative-parity partner of the nucleon propagating
backward in time we employ the Dirichlet boundary condition in the temporal
direction for quark propagators. We used the relativistic nucleon operator
$N=({}^{t}qC^{-1}\gamma_{5}q)q$.  Errors are estimated by the single
elimination jackknife procedure.

In Fig.~2 the ratio $R(t)$ is plotted for the connected and
disconnected  amplitudes as a function
of $t$ for the case of $K=0.160$ and 0.1665.  For the connected
amplitude we observe a clear linear increase
with very small errors of less than 1--2\% up to $t\approx 10$  as has been
known from previous work\cite{threepoint,gupta}.  Signals are also
reasonable, albeit worse, for the disconnected amplitude, and are consistent
with
a linear behavior in $t$ up to $t\approx 10$.  The slope of the disconnected
amplitude is even larger than that of the connected one, indicating a
substantial contribution from the disconnected amplitude to the $\sigma$ term.
To extract the scalar density matrix element we fit the  data to the linear
form
({\ref{eq:ratio}}) with the fitting range chosen to be $4\leq t\leq 9$. In
Table~1 we tabulate the fitted values of the matrix elements corrected by the
tadpole-improved renormalization factor in the ${\overline{MS}}$ scheme at the
scale $\mu=1/a$  given by{\cite{tadpole}} \begin{equation}
Z_S=\left(1+\alpha_s\left(\frac{2}{\pi}\log a\mu-0.0098\right)\right)
\left(1-\frac{3K}{4K_c}\right) ,
\label{eq:zfactor}
\end{equation}
where we used $\alpha_{\overline{MS}}(1/a)=0.2207$ for $\alpha_s$.

We present the quark mass dependence of the matrix element
$<N|{\bar u}u+{\bar d}d|N>$ in Fig.~3 as functions of the
bare quark mass $m_q=(1/K-1/K_c)/2$ using $K_c=0.1694${\cite
{GF11}}. The disconnected contribution (circles) is about twice as large as
the connected contribution and increases almost linearly toward small quark
masses, while the connected contribution remains constant.  Extrapolating
linearly in the quark mass, we find $<N|{\bar u}u+{\bar d}d|N>=5.8(1.4)$ for
the
disconnected contribution, 2.62(6) for the connected piece, and 8.6(1.4) for
the sum at $m_q=0$.  The value for the connected contribution is
reasonably consistent with the value of the derivative $dm_N/dm_q=2.97(11)$
obtained by fits of the nucleon mass in the range $K=0.160-0.1665$ where our
calculations are made.

An estimate of the $\sigma$ term in physical units requires the value of the
bare mass\cite{quarkmassfactor} $\hat m=(m_u+m_d)/2$ for the up and down quark
and the physical scale of lattice spacing.  From fits of the meson spectrum
obtained on our set of gauge configurations, we find $m_\pi^2=2.710(32)\hat
m$,
$m_\rho=0.5294(86)+1.593(61)\hat m$, from which we obtain $\hat m=0.00342(12)$
and $a^{-1}=1.45(2)$GeV using $m_\pi=140$MeV and $m_\rho=770$MeV.  These
values
yield $\sigma=43(7)$MeV, showing an encouraging agreement with the dispersion
theoretic estimate of the $\sigma$ term.  Alternatively one can form the
dimensionless ratio $m_N\sigma/m_\pi^2$ at each quark mass and extrapolate to
the chiral limit.  The ratio turned out to depend little on $m_q$, and a
linear
extrapolation gives $m_N\sigma/m_\pi^2=2.80(53)$ or $\sigma=58(11)$MeV with
the
physical nucleon and pions masses.  The difference between the two values of
$\sigma$ originates from the fact that $m_N/m_\rho=1.48(4)$  at $\beta=5.7$,
taken for the present simulation, is larger than the experimental value 1.22.

Let us turn our attention to the nucleon matrix element of the
strange quark density $<N|{\bar s}s|N>$. This matrix element receives
contribution only from the disconnected amplitude.  For an extraction of its
physical value one needs to extrapolate the valence quark mass to the
chiral limit keeping the strange quark mass fixed.  In Table~2 we summarize
the
results for the matrix element $<N|{\bar s}s|N>$ corrected by
the tadpole-improved $Z$ factor relevant for such an analysis.
Results of a linear extrapolation to the chiral limit for the valence quark
are
also listed in Table~2. Fitting the extrapolated values for three values of
$K_s$ we obtain   $<N|{\bar s}s|N>_{K_{val}=K_c}=2.93(93)-1.1(7.9)m_s$.

The strange quark mass may be estimated by generalizing the relation
$m_\pi^2=2.710(32)\hat m$ to $m_K^2=2.710(32)(\hat m+m_s)/2$ and using the
experimental ratio $m_K/m_\rho=0.64$, which yields $m_s=0.0831(30)$ or
$K_s=0.1648$. We then obtain $<N|{\bar s}s|N>=2.84(44)$.  Two quantities of
physical interest that can be estimated from this result are the strange
quark contribution to the nucleon mass $m_s<N|{\bar
s}s|N>/m_N=0.302(48)$ and the $K$-$N$ $\sigma$ term
$\sigma_{KN}=(\hat m+m_s)$ $<N|\bar uu+\bar dd+2\bar
ss|N>/4=0.310(37)=451(54)$MeV.   While these results appear quite reasonable,
we
need to note that the value of the matrix element $<N|{\bar s}s|N>$ itself is
large.  Combined with our result  $<N|{\bar u}u+{\bar
d}d|N>=8.6(1.4)$ for the up and down quark, we find  $y=2<N|{\bar
s}s|N>/<N|{\bar u}u+{\bar d}d|N>=0.66(15)$, while a phenomenological
estimate gives $y\approx 0.2$\cite{update}.  We have not identified an
apparent origins of systematic errors that lead to the result, perhaps except
for a possible scaling violation as our simulation is made at a rather strong
coupling.  This point should be examined through a repetition of the
calculation with a
smaller lattice spacing, which we leave for future investigations.

To summarize we have shown that the method
of wall source without gauge fixing\cite{pipisct,etamass} helps to overcome
the
computational difficulty of yet another quantity, the disconnected
contribution to the $\pi$-$N$ $\sigma$ term, for which we obtained
$\sigma=40-60$MeV with a 15\% error at a lattice spacing $a\approx 0.14$fm.
Our
results also show a large strangeness content in the nucleon.

\section*{Acknowledgements}

Numerical calculations for the present
work have  been carried out on HITAC S820/80 at KEK.  One of us (AU) thanks T.
Hatsuda for useful discussions.  This work is supported in part by the
Grants-in-Aid of
the Ministry of Education (Nos. 05NP0601, 05640325, 05640363, 05-7511,
06640372).

\newpage
\begin{center}

\end{center}

\newpage
\begin{center}
\section*{Figure Captions}
\end{center}

\begin{itemize}
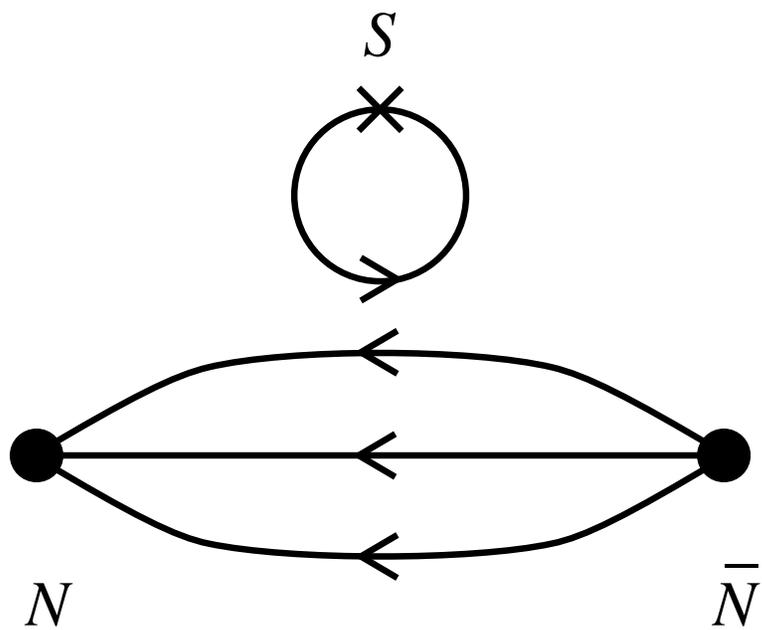
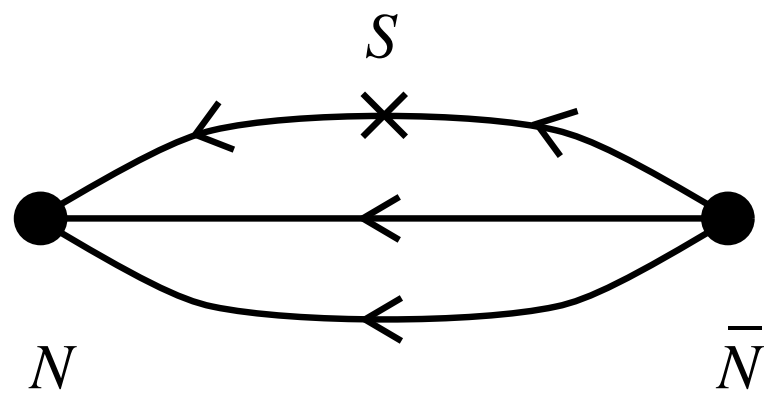
\item[Fig.~1]  (a) Disconnected and (b) connected contributions to the
nucleon-scalar density three-point function.

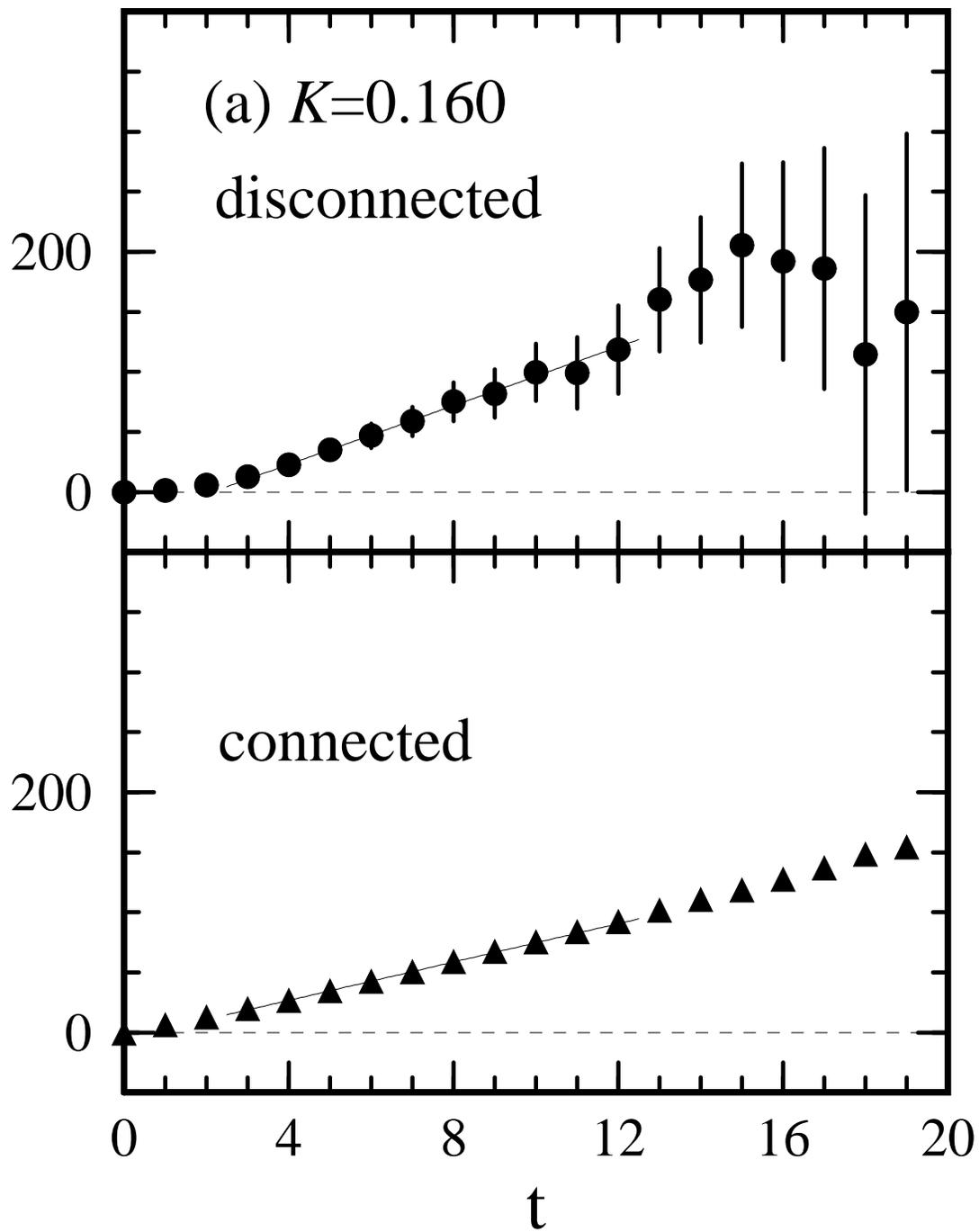
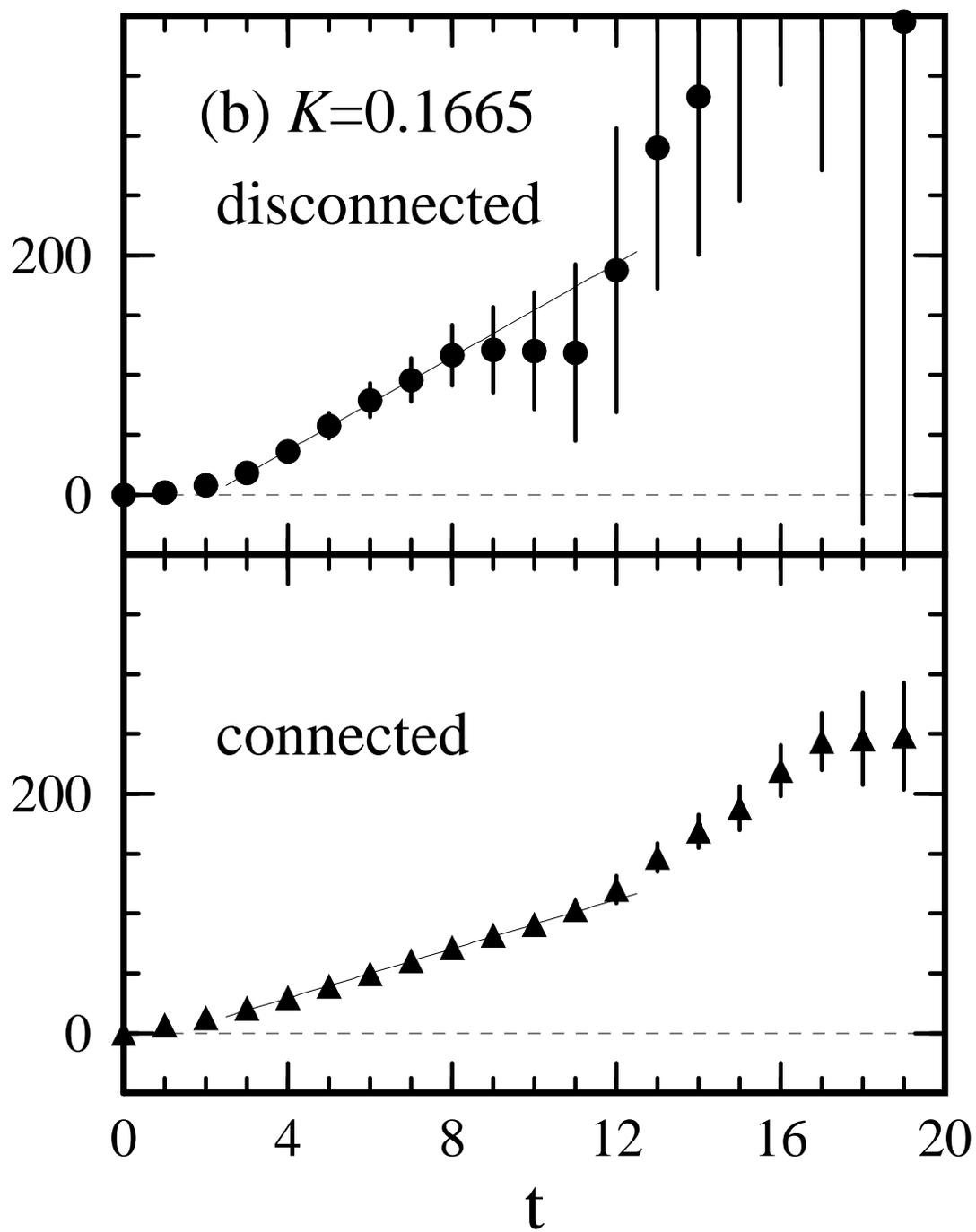
\item[Fig.~2] Ratio $R(t)$ for the disconnected (circles) and
connected (triangles) amplitudes  at $\beta=5.7$ on a $12^3\times 20$
lattice. Solid lines are linear fits over $4\leq t\leq 9$. (a) $K=0.160$ and
(b) $K=0.1665$.

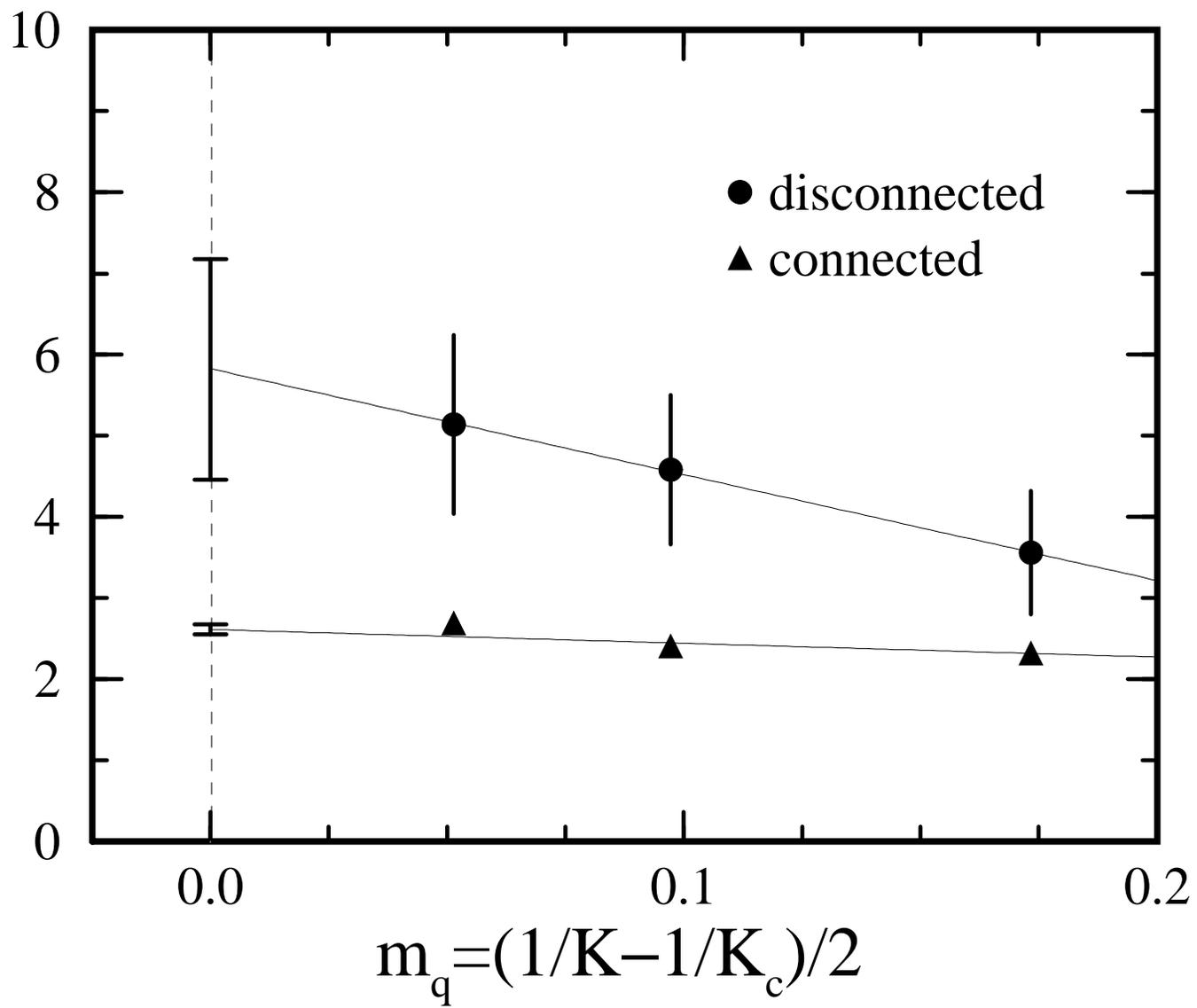
\item[Fig.~3] Disconnected (circles) and connected (triangles)
 contribution to the matrix element $<N|{\bar
u}u+{\bar d}d|N>$ as a function of the quark mass $m_q=(1/K-1/K_c)/2$ in
lattice units.  Solid lines are linear fits for extrapolation to the chiral
limit
$m_q=0$.
\end{itemize}

\newpage
\begin{center}
\section*{Tables}
\end{center}

\begin{table}[h]
\begin{center}
\begin{tabular}{cccc|c}
\hline
$K$                                     & 0.160 & 0.164 & 0.1665 &
$K_c=0.1694$ \\
\hline
$m_q$                                   & 0.173 & 0.097 & 0.051  &          \\
\# conf.                           & 300   &  300  &  400   & \\
$m_{\pi}/m_{\rho}$                      & 0.85  & 0.74  &  0.60  &  \\
$m_N$
&1.2957(79)&1.080(10)&0.926(13)&0.783(16)\\
$<N|{\bar u}u+{\bar d}d|N>_{\rm conn.}$
&2.323(15)&2.413(30)&2.693(81)&2.615(61)\\
$<N|{\bar u}u+{\bar d}d|N>_{\rm disc.}$
&3.56(76)&4.58(92)&5.14(1.10)&5.8(1.4)\\
\hline
\end{tabular}
\caption{Disconnected and connected contributions to the
$\sigma$ term matrix element $<N|{\bar u}u+{\bar d}d|N>$  at $\beta=5.7$ on a
$12^3 \times 20$ lattice in quenched QCD.  Results are corrected by the
tadpole-improved $Z$ factor.  Values at $K_c$ are obtained by a linear fit in
$1/K$.}
\end{center}
\end{table}

\begin{table}[h]
\begin{center}
\begin{tabular}{cccccc}
\hline
&                & \multicolumn{2}{c}{$<N|{\bar s}s|N>$} & \\
& $K_{\rm val.}$ & $K_s=0.160$ & $K_s=0.164$ & $K_s=0.1665$    \\
\hline
&  0.160    & 1.78(38) & 2.02(38) & 2.20(39) \\
&  0.164    & 2.04(47) & 2.29(46) & 2.49(45) \\
&  0.1665   & 2.50(55) & 2.81(55) & 2.57(55) \\
\hline
&  $K_c=0.1694$ & 2.67(69) & 2.99(68) & 2.77(68) \\
\hline
\end{tabular}
\caption{$<N|{\bar s}s|N>$ as a function of the hopping parameters of valence
and strange quarks obtained at $\beta=5.7$ on a $12^3 \times
20$ lattice  in quenched QCD with  300  gauge configurations..   Results are
corrected by the tadpole-improved $Z$ factor. }
\end{center}
\end{table}


\begin{thebibliography}{99}
\bibitem{quarkmassfactor}The renormalization factor for the quark mass equals
the inverse of that for the scalar density to all orders in perturbation
theory, and hence the two factors cancels out in the $\sigma$ term.  This
point
is taken into account in our estimate of $\sigma$ below.
\end{thebibliography}
\end{document}